\documentclass[proof]{WileyASNA-v1}
\usepackage{color}

\articletype{Comment}%

\received{23 December 2024}
\revised{}
\accepted{}

\def\aap{A\&A}
\def\apjl{ApJ}

\def\apjs{ApJS}

\def\apjs{ApJ Supplements}

\def\apjl{ApJ Letters}

\def\aap{A\&A}

\begin{document}

\title{The lens mass in the Einstein ring JWST-ER1}

\author[1]{Fulvio Melia*}

\authormark{Fulvio Melia}

\address[1]{\orgdiv{Departments of Physics and Astronomy, and the Applied Math Program}, 
\orgname{The University of Arizona}, \orgaddress{\state{Tucson, Arizona}, \country{U.S.A.}}}

\corres{*\email{fmelia@arizona.edu}}


\abstract{JWST has made several surprising discoveries, underscored by the `too early' 
appearance of well-formed galaxies and supermassive black holes. It recently also 
uncovered a compact galaxy (JWST-ER1g) associated with a complete Einstein ring 
(JWST-ER1r) at photometric redshift $z_l=1.94^{+0.13}_{-0.17}$, produced by a
lensed galaxy at $z_s=2.98^{+0.42}_{-0.47}$. In two independent studies, this 
system ($M_{\rm lens}\sim 6.5\times 10^{11}\;M_\odot$) has yielded different 
conclusions concerning whether or not it requires an unexpected contribution 
of mass from sources other than stars and fiducial dark matter. The different 
redshift inferred by these two analysis for the lensed galaxy appears to be the 
culprit. In this paper, we examine the impact of the background cosmology on our 
interpretation of the JWST data. We compare the measured characteristics of 
JWST-ER1 in flat-$\Lambda$CDM with those emerging in the context of 
$R_{\rm h}=ct$. We show that, unlike the latter model's mitigation of the 
tension created by JWST's other discoveries, neither cosmology is favored 
by this particular Einstein ring. The disparity is common to both models, 
leaving open the question of whether a new type of mass, or a modified initial 
mass function, may be present in this source.}


\maketitle

\section{Introduction}\label{intro}
The Einstein ring, dubbed JWST-ER1r by \cite{vanDokkum:2024}, and the lensing galaxy 
(JWST-ER1g) that created it, were identified in James Webb Space Telescope (JWST) NIRCam 
observations, as part of the COSMOS-Web project \citep{Casey:2023}. From this study, one
infers that the lens is a compact, massive, quiescent galaxy at a photometric redshift of 
$z_l=1.94^{+0.13}_{-0.17}$, and the ring is produced by a background galaxy at 
$z_s=2.98^{+0.42}_{-0.47}$, with a radius of $0.77''\pm0.01''$, corresponding to a scale 
of $r_{\rm ER}=6.6$ kpc in the context of $\Lambda$CDM. The ring is large enough to provide 
a direct measurement of the central mass of a large galaxy before the subsequent 10 Gyr of 
evolution would have mixed and diluted its stellar content.

A second, independent discovery of this system \citep{Mercier:2023}, known alternatively
as the COSMOS-Web ring, confirms many of these features, except in one key respect---the
photometric redshift of the lensed galaxy. They report a photometric redshift of $2.02\pm 0.02$
for the lens, an Einstein ring radius of $0.78''\pm0.04''$, and a photometric redshift of
$5.48\pm 0.06$ for the source. As we shall see shortly, this disparity significantly
impacts the inferred properties of the lens.

The modelling of JWST-ER1g is facilitated by the fact that it is almost perfectly
round, with no obvious star-forming regions, tidal tails, or any other irregularities.
Its intensity distribution is well fit with a S\`ersic model \citep{Sersic:1968} (see
Eq.~\ref{eq:Sersic} below), with a S\`ersic index $n=5.0\pm0.6$ and an effective radius 
of the galaxy $r_e=0.22''\pm0.02''$, corresponding to a standard model size of $1.9$ kpc, 
well within the radius of the ring \citep{vanDokkum:2024}. These features change somewhat
in the second study, but are perfectly conistent with each other within the quoted errors.

In the analysis of \cite{vanDokkum:2024}, a possible `missing mass' problem begins to emerge 
from the Prospector \citep{Johnson:2021} total stellar mass of this galaxy configuration, 
assuming a Chabrier initial mass function (IMF) \citep{Chabrier:2003}, which is calculated 
to be $M_{\rm gal}=1.3^{+0.3}_{-0.4}\times 10^{11}\;M_\odot$, of which 
$M_*=1.1^{+0.2}_{-0.3}\times 10^{11}\; M_\odot$ is contained within $r_{\rm ER}$.

This is because the photometric redshifts of the source and lens, combined with the measured 
value of the ring's radius, give a total lens mass $M_{\rm lens}=6.5^{+3.7}_{-1.5}\times
10^{11}\;M_\odot$ (see Eq.~\ref{eq:rER} below). In their exhaustive analysis of other
possible contributors to $M_{\rm lens}$, \cite{vanDokkum:2024} rule out all but one
likely source---the projected dark matter mass within the ring from the halo hosting 
the lens galaxy JWST-ER1g. They find a total dark matter mass $M_{\rm dm}=2.6^{+1.6}_{-0.7}
\times 10^{11}\;M_\odot$ within $r_{\rm ER}$, which falls short by about $2.8\times 10^{11}
\;M_\odot$, leading to the conclusion that some other, yet unknown, source must be 
contributing to the overall lens mass in this system.

\begin{figure*}
\centering
\includegraphics[width=5.5in]{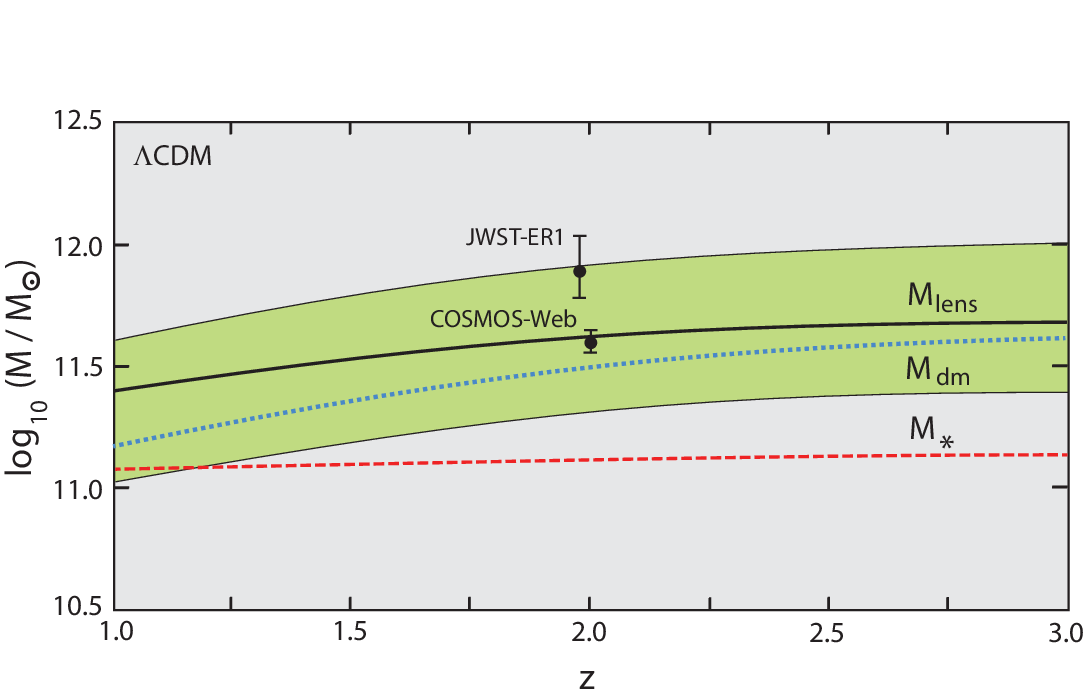}
	\caption{The total lens mass (black solid) as a function of $z$ for a galaxy with the 
	intensity profile of JWST-ER1g. The green shaded region represents the $1\sigma$ uncertainty, 
	estimated from the errors in the total stellar mass of JWST-ER1g and the total halo mass 
	inferred from the dark matter-stellar mass relation. The two dominant contributions, stellar 
	(red dashed) assuming a Chabrier initial mass function, and dark matter (blue dotted), 
	correspond to the integrated mass within the observed JWST-ER1r Einstein ring, whose radius 
	($6.6$ kpc) at $z=1.94$ and source redshift $z_s=2.98$ implies a lensing mass 
	$M_{\rm lens}=6.5^{+3.5}_{-1.5}\times 10^{11}\;M_\odot$, as shown by the data point 
	labeled `JWST-ER1' and its error bar. By comparison, the lens mass is inferred to be 
	$M_{\rm lens}=(3.66\pm0.36)\times 10^{11}\;M_\odot$ if $z_s=5.48$ (as indicated by the 
	point labeled `COSMOS-Web'). The principal cause of the variation in $M_{\rm dm}$ with 
	$z$ is the redshift dependence of the critical density, $\rho_{\rm c}(z)$, which is model 
	dependent. The plots in this figure confirm the conclusions drawn by 
	\cite{vanDokkum:2024} and \cite{Mercier:2023} in the context of 
	flat-$\Lambda$CDM, i.e., that $M_*$ and $M_{\rm dm}$ fully account for $M_{\rm lens}$
	as long as $z_s=5.48$, but not for $z_s=2.98$.}
\label{fig1}
\end{figure*}

Interestingly, \cite{Mercier:2023} come to very similar conclusions concerning $M_*$ and 
$M_{\rm dm}$, but differ considerably on whether these are sufficient to account for the 
lens mass, which they estimate to be $(3.66\pm0.036)\times 10^{11}\;M_\odot$ from 
Equation~(\ref{eq:rER}), given their different source redshift. Thus, unlike \cite{vanDokkum:2024},
they find that the lens in the COSMOS-Web ring does not require an additional, unknown source of 
mass.

Motivated by other recent model comparisons in the context of JWST's remarkable sequence
of surprising discoveries \citep{Melia:2023b,Melia:2024c}, we here extend this interesting
work by examining how---if at all---our conclusions regarding the lens structure in JWST-ER1
may be influenced by the background cosmology. We thus carry out the analysis for two 
competing models, including the $R_{\rm h}=ct$ cosmology \citep{Melia:2007,MeliaShevchuk:2012,Melia:2020}, 
in addition to $\Lambda$CDM. As we shall see very shortly, the internal structure of the dark-matter
halos depends critically on their formation and history, notably via the well-studied 
mass-concentration-redshift relation \citep{Ludlow:2014}. As such, it would not suprise us
to learn that the interplay between $M_*$ and $M_{\rm dm}$ in forming the lens may be
self-consistent in some evolutionary scenarios and not in others. The redshift-dependence
of $M_{\rm halo}$ for the same $M_*$ will thus feature prominently in our analysis.

\section{Lens mass}\label{lensmass}
In a spatially flat Universe, the mass of the lens within $r_{\rm ER}$ is given by
\begin{equation}
	M_{\rm lens}(<\theta)={\theta^2c^2\, d_A^{\,l}\,d_A^{\,s}\over 4G\,d_A^{\,ls}}\;,\label{eq:rER}
\end{equation}
where $\theta$ is the measured Einstein radius in radians, $d_A^{\,l}$ is the model-dependent
angular diameter distance to the lens, $d_A^{\,s}$ is the angular diameter distance to the source, and 
\begin{equation}
	d_A^{\,ls}\equiv d_A^{\,s}-{1+z_l\over 1+z_s}d_A^{\,l}\label{eq:dls}
\end{equation}
is the angular diameter distance between the source and lens. The other symbols have their
usual meanings. 

In flat $\Lambda$CDM, we have 
\begin{equation}
	d_{A,\,\Lambda}^{\,ls}={c\over H_0}{1\over 1+z_s}\int_{z_l}^{z_s}{du\over
	\sqrt{\Omega_{\rm m}(1+u)^3+\Omega_{\rm r}(1+u)^4+\Omega_\Lambda}}\;,\label{eq:dAlcdm}
\end{equation}
where $\Omega_{\rm m}$, $\Omega_{\rm r}$ and $\Omega_\Lambda$ are, respectively, the matter,
radiation and dark energy densities scaled to $\rho_{\rm c}(0)$ today. For the principal result
discussed in this paper, there is no need to optimize the cosmological parameters,
so we simply assume the fiducial values $H_0=70$ km s$^{-1}$ Mpc$^{-1}$, $\Omega_{\rm m}=0.3$
and $\Omega_\Lambda=1-\Omega_{\rm m}-\Omega_{\rm r}$. Given the measured angular size of
JWST-ER1r, we thus infer a spatial scale $r_{{\rm ER},\,\Lambda}\approx 6.6$ kpc in this model.
The measurements of \cite{Mercier:2023} are fully consistent with these results. 

Correspondingly, the angular diameter distance in $R_{\rm h}=ct$ is given as \citep{Melia:2020}
\begin{equation}
	d_{A,\,{R_{\rm h}}}^{\,ls}={c\over H_0}{1\over 1+z_s}
	\ln\left({1+z_s\over 1+z_l}\right)\label{eq:dARh}
\end{equation}
and, for simplicity, we simply adopt the same Hubble constant here. This model
has no other free parameters. In this scenario, we find an Einstein ring radius
$r_{{\rm ER},\,R_{\rm h}}\approx 6.0$ kpc.

The ring is so large compared to JWST-ER1g's effective radius that most of the stellar mass is 
contained within it. The determination of the lensing galaxy's mass is based on well understood 
and tested astrophysical principles, leaving little room for doubt concerning its contribution 
to $M_{\rm lens}$. The S\`ercic  model for its intensity profile, 
\begin{equation}
	I(r) = I_e\exp\left\{-b_n\left[\left({r\over r_e}\right)^{1/n}-1\right]\right\}
	\;,\label{eq:Sersic}
\end{equation}
where $r_e$ is the half-light radius and $I_e$ is the intensity at that radius, and
$b_n\approx 2n-1/3$ in terms of the S\`ersic index $n$, fits the five-band photometry 
of JWST-ER1g very well \citep{vanDokkum:2024}. Indeed, the lensing galaxy falls comfortably
on the canonical size-mass relation of quiescent galaxies at $z\sim 2$ \citep{vanderWel:2014}.

A caveat to this viewpoint, however, concerns the adopted IMF, which is conventionally 
assumed to be the Chabrier distribution. If some of the mass is missing, perhaps it is in the 
form of an overabundance of low mass stars, and \cite{vanDokkum:2024} do show that
bottom-heavy IMFs could in principle account for most of $M_{\rm lens}$ in such a scenario.
But such unorthodox IMFs appear to be inconsistent with the comparison of dynamical
to stellar masses in $z\sim 2$ galaxies, which instead favor bottom-light
IMFs, such as the Chabrier form \citep{Esdaile:2021}.

\begin{figure*}
\centering
\includegraphics[width=5.5in]{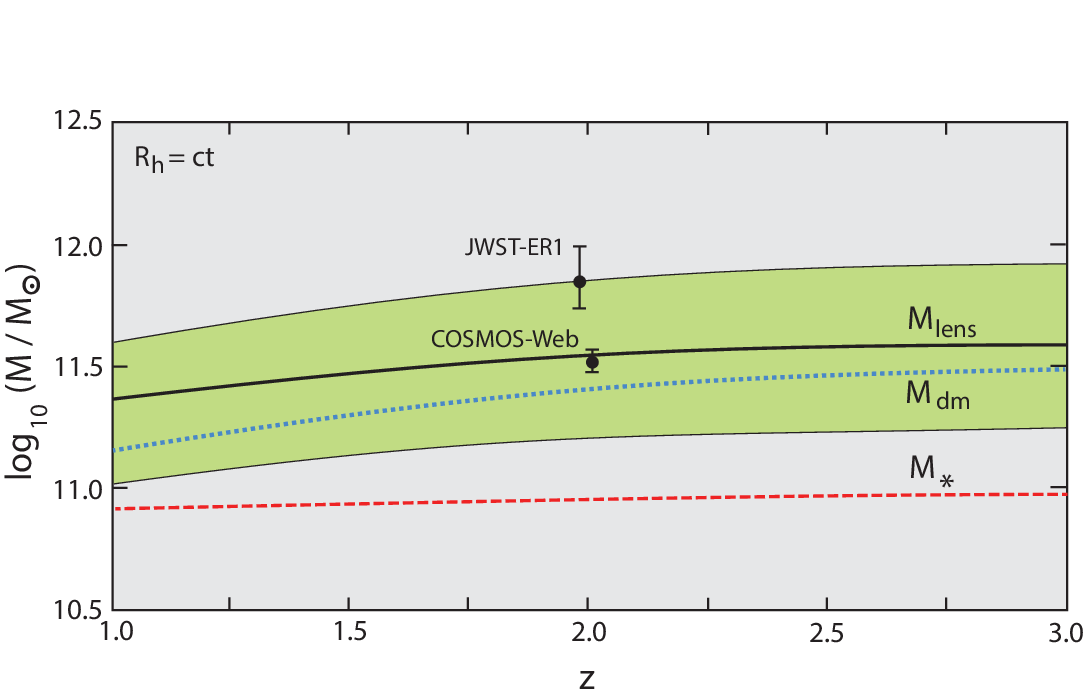}
	\caption{Same as Fig.~\ref{fig1}, except now for the $R_{\rm h}=ct$ model. The green
	shaded region has the same meaning as in this earlier figure. Though the geometry of 
	the Universe is different in these two cosmologies, e.g., as inferred from the 
	angular diameter distances, the comparison between the measured 
	($5.3^{+3.0}_{-1.3}\times 10^{11}\;M_\odot$ for $z_s=2.98$; and 
	$3.0\pm0.3\times 10^{11}\;M_\odot$ for $z_s=5.48$) and calculated lens masses 
	follows the same pattern in these two scenarios. Specifically, the COSMOS-Web analysis
	is fully consistent with $M_*$ and $M_{\rm dm}$, while the JWST-ER1 result
	appears to signal missing mass in the lens, independently of the background cosmology.}
\label{fig2}
\end{figure*}

To make our discussion concerning the impact of a redshift-dependent structure of the 
dark matter halo on the interpretation of $M_{\rm lens}$ versus $M_*$ in JWST-ER1
as transparent as possible, we here take the minimalist approach of simply assuming
the conventional stellar mass distribution consistent with Equation~(\ref{eq:Sersic})
and a Chabrier IMF. Moreover, we examine the redshift variation of $M_{\rm lens}$
expected over the range $1\le z\le 3$ for a galaxy with the same mass profile as that
inferred for JWST-ER1g under these conditions. This galaxy's contribution to the 
lensing mass is indicated by the red, dashed curve in Figures~\ref{fig1} and
\ref{fig2}. The slow variation with redshift is due to the changing fraction of 
its total stellar mass contained within the Einstein ring, whose spatial scale also 
changes with $z$.

A much bigger variation with $z$ is expected for $M_{\rm dm}$, given that the dark matter
halo extends over a much larger size than $r_{\rm ER}$, so that the fraction contained within
the ring is more susceptible to a change in the spatial scale. The total mass in the halo,
$M_{\rm halo}$, is constrained by the observed halo mass-stellar mass relation \citep{Behroozi:2013} 
which, for $0\le z\le 3$, falls within the range $10^{13}\;M_\odot\lesssim M_{\rm halo}
\lesssim 10^{15}\;M_\odot$ for $10^{11}\;M_\odot\lesssim M_{\rm gal}\lesssim 2\times 10^{11}\;M_\odot$.
At $M_{\rm gal}=1.3^{+0.3}_{-0.4}\times 10^{11}\;M_\odot$, one finds $M_{\rm halo}=1.0^{+2.6}_{-0.5}
\times 10^{13}\;M_\odot$. The internal structure of this halo, however, including the
concentration of its mass towards the center and the average density of dark matter within
its virial radius, varies with redshift, a key point of this paper. 

The projected dark matter mass within the ring, $M_{\rm dm}$, may be estimated by integrating
a standard NFW profile \citep{Navarro:1997} in a cylinder along the line-of-sight with a 
radius $r_{\rm ER}$ \citep{Lokas:2001}. The equilibrium density profile of the halo is now well 
established, and may be approximated by the scaling relation
\begin{equation}
	\rho(r)={\delta_{\rm c}\rho_{\rm c}(z)\over (r/r_s)(1+r/r_s)^2}\;,\label{eq:NFW}
\end{equation}
characterized by the scale radius $r_s$, overdensity $\delta_{\rm c}$ and critical density
$\rho_{\rm c}(z)$ at the redshift where the halo is observed \citep{Ludlow:2014}. The halo's
profile may also be expressed in terms of a dimensionless `concentration' parameter, $c\equiv
r_{200}/r_s$, the ratio of its virial to scale radii, and its virial mass, $M_{200}$, 
contained within a sphere enclosing a mean density $200$ times the critical density,
$\rho_{\rm c}(z)\equiv 3H(z)^2/8\pi G$. For example, one may write
\begin{equation}
	\delta_{\rm c}={200\over 3}{c^3\over \ln(1+c)-c/(1+c)}\;.\label{eq:c}
\end{equation}

It has been known since the early work of \cite{Navarro:1997} that the concentration (and
therefore the average density of dark matter) and mass are strongly correlated, and 
considerable effort has been made to understand the
origin of this coupling and its dependence on redshift. One may easily gauge the impact 
of redshift on $c(z)$ from an inspection of Figure~1 in \cite{Ludlow:2014}, which shows 
the variation of the concentration parameter as a function of halo mass over the redshift 
range $0\le z\le 2$. For example, at $M_{200}\sim 10^{13}\;M_\odot$, $c(z)$ drops from
$\sim 7$ at $z=0$ to $\sim 4$ at $z=2$, a variation that significantly affects the density
profile in Equation~(\ref{eq:NFW}) and thereby the fraction of $M_{\rm halo}$ contained
within $r_{\rm ER}$.

The blue dotted curves in Figures~\ref{fig1} and \ref{fig2} show the dark matter enclosed 
within the Einstein ring as a function of redshift, under the assumption that: (i) the
lensing galaxy's total stellar mass is held fixed at the value inferred for JWST-ER1g;
(ii) the halo mass, $M_{\rm halo}$ is extracted from Figure~7 in \cite{Behroozi:2013};
and (iii) $\rho_{\rm c}(z)$ in Equation~(\ref{eq:NFW}) has the redshift dependence 
predicted by $\Lambda$CDM (Fig.~\ref{fig1}),
\begin{equation}
	\rho_{\rm c}^\Lambda(z)=\rho_{\rm c}(0)\left[\Omega_{\rm m}(1+z)^3
	+\Omega_{\rm r}(1+z)^4+\Omega_\Lambda\right]\;,\label{eq:rhocL}
\end{equation}
or $R_{\rm h}=ct$ (Fig.~\ref{fig2}),
\begin{equation}
	\rho_{\rm c}^{R_{\rm h}}(z)=\rho_{\rm c}(0)(1+z)^2\;.\label{eq:rhocR}
\end{equation}

\begin{figure*}
\centering
\includegraphics[width=5.5in]{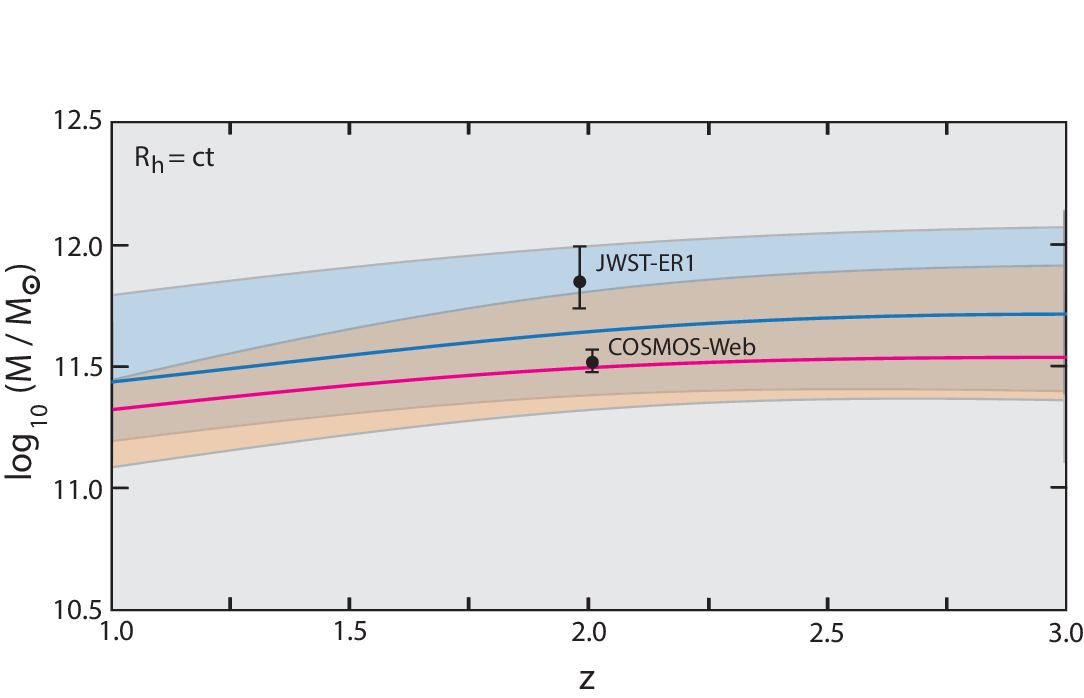}
	\caption{Same as Fig.~\ref{fig2}, except now showing the lens mass in 
	$R_{\rm h}=ct$ with a concentration parameter, $c$, smaller (red) and greater (blue) by 
	$25\%$ than the fiducial value used in Fig.~\ref{fig2}. The shaded regions have the 
	same meaning as in the earlier figures. The larger concentration parameter brings the 
	theoretical prediction somewhat closer to the JWST-ER1 result, but the lens masses 
	overall follow the same pattern as in the previous two scenarios. Specifically, the 
	COSMOS-Web analysis remains fully consistent with the sum of $M_*$ and $M_{\rm dm}$, 
	while the JWST-ER1 result is in tension with the calculated lens mass, independently 
	of the background cosmology.}
\label{fig3}
\end{figure*}

The solid black curves in these figures then show the mass $M_*+M_{\rm dm}$,
along with the $1\sigma$ uncertainty (green) estimated from the errors in the total
stellar mass of JWST-ER1g and the total halo mass inferred from the dark matter-stellar
mass relation. An important conclusion from these plots is that the fraction of 
$M_{\rm halo}$ contained within the Einstein ring changes noticeably as a function 
of redshift. Even so, the differences between flat-$\Lambda$CDM and $R_{\rm h}=ct$
are not sufficient to leverage this variation into an explanation for the missing
mass in JWST-ER1. We find a total mass $M_*+M_{\rm dm}=3.2^{+3.1}_{-2.0}\times 10^{11}\;M_\odot$
in $R_{\rm h}=ct$, which matches the measured lens mass, $M_{\rm lens}=(3.0\pm 0.3)
\times 10^{11}\;M_\odot$, very well for $z_s=5.48$, but not for $z_s=2.98$, where
the lens mass is instead $5.3^{+3.0}_{-1.3}\times 10^{11}\;M_\odot$.
\vfill

\section{Conclusion}\label{conclusion}
A possible caveat to this conclusion is that, although perturbation growth 
calculations have been carried out in $R_{\rm h}=ct$, no detailed N-body simulations
comparable to the extensive work completed for $\Lambda$CDM have been done yet. So a study 
of $c(M,z)$ based on the mass-accregion history in the former cosmology is not yet feasible. 
Having said this, there are good reasons to believe that, once the initial halo is formed, 
its subsequent growth history should not be too different between $\Lambda$CDM and 
$R_{\rm h}=ct$. This is indicated by the Birkhoff theorem \citep{Melia:2020}, which ensures 
that the spacetime within the halo's boundary is independent of what is happening in the 
exterior region. In other words, the expansion history of the Universe does not affect the 
internal growth of the halo once it has become gravitationally bound.

As such, the functional form of the concentration profile described in \cite{Ludlow:2014}
and \cite{Ludlow:2016} should be quite similar in $\Lambda$CDM and $R_{\rm h}=ct$. What 
varies between these two models is the scaling. These previous studies suggest that,
at $M_{\rm halo}\sim 10^{13}\;M_\odot$ and $z\sim 2$ (as we have here), the largest 
variation one can expect in $c(M,z)$ among the various cosmologies being tested is 
about $25\%$. The differences are larger for smaller halos, but these are not relevant 
to this work.

Thus, we may bracket the uncertainty in $c(M,z)$ for $R_{\rm h}=ct$ relative
to $\Lambda$CDM by comparing the results of our simulations for a range of
concentrations between $0.75$ and $1.25$ times the fiducial value of $c(M,z)$. 
The corresponding results are plotted in Figure~\ref{fig3}, which shows the calculated
lens mass in $R_{\rm h}=ct$ for the larger concentration (blue) versus the smaller
one (red). The larger concentration certainly renders the theoretical prediction more
in line with the JSWT-ER1 result, but the overall pattern is unchanged from the
scenarios shown in Figures~\ref{fig1} and \ref{fig2}. That is, while the COSMOS-Web
analysis remains fully consistent with the sum of $M_*$ and $M_{\rm dm}$, the 
JWST-ER1 result is in tension with the calculated lens mass, independently of 
the background cosmology.

In the scant two years of operation, JWST has already provided a multitude of surprises 
that indicate a need for reavaluating our current understanding of the early Universe 
\citep{Melia:2023b,Melia:2024c}. The most recent discovery of a complete Einstein ring
associated with the compact galaxy JWST-ER1g seemingly creates yet another 
anomaly, this time requiring either a significant modification to the initial mass function 
at $z\sim 2$ or, more provocatively, a new source of mass for the lens in this system, 
beyond stellar and fiducial dark matter. Attempts have also been made to refine the
behavior of dark matter in these systems, by postulating a restructuring of the
halo due to self-interactions \citep{Kong:2024}. 

As we have seen, the missing mass problem goes away if the source redshift is
$z_s=5.48$ instead of $2.98$. In this paper, we have extended the previous work
of \cite{vanDokkum:2024} and \cite{Mercier:2023} to see what impact, if any,
the choice of cosmology may have on this possible anomaly. But though the geometry
of the Universe is somewhat different in these two models, the redshift-dependence
of the dark-matter halo's internal structure is not sufficient to resolve
the problem for $z_s=2.98$. 

Unlike JWST's other discoveries, the measured properties of the JWST-ER1/COSMO-Web ring 
therefore do not appear to favor either of these two models.

\section*{Acknowledgments}
I am grateful to the anonymous referee for helpful comments that have improved
the presentation of this material. 

\subsection*{Author contributions}

Fulvio Melia is the sole author of this paper.

\subsection*{Financial disclosure}

None reported.

\subsection*{Conflict of interest}

The authors declare no potential conflict of interests.

\bibliography{ms}%

\end{document}